\documentclass[twocolumn,showpacs,preprintnumbers,amsmath,amssymb]{revtex4}

\usepackage{graphicx}
\usepackage{dcolumn}
\usepackage{bm}

\begin{document}

\title{Thermodynamic properties of a dipolar Fermi gas}
\author{J.-N. Zhang and S. Yi}

\affiliation{Key Laboratory of Frontiers in Theoretical Physics, Institute of Theoretical Physics, Chinese Academy of Sciences, Beijing 100190, China}

\begin{abstract}
Based on the semiclassical theory, we investigate the thermodynamic properties of a dipolar Fermi gas. Through a self-consistent procedure, we numerically obtain the phase-space distribution function at finite temperature. We show that the deformations in both momentum and real space become smaller and smaller as the temperature is increased. For the homogeneous case, we also calculate pressure, entropy, and heat capacity. In particular, at the low-temperature limit and in the weak interaction regime, we obtain an analytic expression for the entropy which agrees qualitatively with our numerical result. The stability of a trapped gas at finite temperature is also explored.
\end{abstract}

\date{\today}

\maketitle

\section{Introduction}
The experimental success in creating ultracold $^{40}$K$^{87}$Rb molecular gas near quantum degeneracy~\cite{ye1,ye2,ye3,ye4} has drawn considerable attention to in studying the fundamental properties of degenerate dipolar Fermi gases. Within the framework of the semi-classical theory, the ground state properties, the collective excitation, and the free expansion dynamics of a normal state dipolar Fermi gas were studied theoretically~\cite{goral,goral2,he}. A recent theoretical work based on a variational approach reveals that, due to the Fock exchange interaction, the momentum distribution is stretched along the direction of dipole moment such that the Fermi surface becomes an ellipsoid~\cite{miya}. This result was confirmed numerically for both homogeneous~\cite{ronen} and trapped~\cite{zhang} systems. Taking into account the effect of the exchange interaction, further theoretical work regarding the normal state of the zero temperature dipolar Fermi gas includes studying the free expansion~\cite{he,sogo}, collective excitation~\cite{goral2,sogo}, zero sound~\cite{ronen}, and the Fermi liquid properties~\cite{chan}.

Another interesting feature of the dipole-dipole interaction is that the partially attractive dipolar force is responsible for the formation of anisotropic BCS pairing~\cite{you,bara,bara2,pu}. With the recent experimental development on control the hyperfine states of $^{40}$K$^{87}$Rb molecules~\cite{ye3,ye4}, the BCS pairings in a mixture of fermionic polar molecules with two different hyperfine states are also studied theoretically~\cite{samo,shi,wu}. In particular, the effects of the Fock exchange interaction to pairing was considered in Ref.~\cite{pu,shi}.

In this paper, we extend our previous work on the ground state properties of the dipolar Fermi gases to finite temperature case. Employing the semi-classical theory, we  numerically obtain the phase-space distribution function through a self-consistent procedure. We show that the deformations of the distribution function in both momentum and real space become smaller and smaller as the temperature is increased. For homogeneous system, we also calculate the thermodynamic quantities such as pressure, entropy, and heat capacity. In particular, at low temperature limit and in weak interaction regime, we derive an analytic expression for the entropy, which agrees qualitatively with our numerical result. For the trapped gases, we also explore the temperature dependence of the stability.

The remainder of this paper is organized as follows. In Sec.~\ref{theory}, we introduce our model and briefly outline the semi-classical theory for a dipolar Fermi gas at finite temperature. The numerical and analytical results are presented in Sec.~\ref{resu}. Finally, we conclude in Sec.~\ref{concl}.

\section{Theory}\label{theory}
Here we consider a system of $N$ spin polarized dipolar fermions interacting via dipole-dipole interaction
\begin{eqnarray}
V_d({\mathbf r})=\frac{c_d}{r^3}\left(1-\frac{3z^2}{r^2}\right),
\end{eqnarray}
where $c_d=d^2/(4\pi\varepsilon_0)$ with $d$ being the electric dipole moment. For simplicity, we have assumed that the dipole moments of all fermions are orientated along $z$-axis. Additionally, we shall consider both homogeneous and trapped systems. In the former case, $U_{\rm ext}$ can be regarded as a box of volume ${\cal V}$; while for the latter, the trap is assumed to be a harmonic potential with axial symmetry, i.e., 
\begin{eqnarray}
U_{\rm ext}({\mathbf r})=\frac{1}{2}m\overline\omega^2\lambda^{-2/3}(x^2+y^2+\lambda^2z^2),
\end{eqnarray}
where $\overline\omega$ is the geometric average of the trap frequencies and $\lambda$ is trap aspect ratio. 

Within the framework of the semiclassical theory, the thermodynamic properties are completely characterized by the phase-space distribution function $f({\mathbf r},{\mathbf k})$, which, at finite temperature $T$, satisfies the Fermi-Dirac statistics
\begin{eqnarray}
f({\mathbf r},{\mathbf k})=\frac{1}{e^{(\varepsilon({\mathbf r},{\mathbf k})-\mu)/k_BT}+1},\label{dist}
\end{eqnarray}
where $\mu$ is the chemical potential introduced to fix the total number of particles such that 
\begin{eqnarray}
\int \frac{d{\mathbf r}d{\mathbf k}}{(2\pi)^3}f({\mathbf r},{\mathbf k})=N.\label{norm}
\end{eqnarray}
The dispersion relation of the quasi-particle takes the form
\begin{eqnarray}
\varepsilon({\mathbf r},{\mathbf
k})&=&\frac{\hbar^2k^2}{2m}+U_{\rm ext}({\mathbf r})+\int\frac{d{\mathbf r}'d{\mathbf k}'}{(2\pi)^3}f({\mathbf r}',{\mathbf k}')V_d({\mathbf r}-{\mathbf r}')\nonumber\\
&&-\int\!\frac{d{\mathbf k}'}{(2\pi)^3} f({\mathbf r},{\mathbf k}')\widetilde
V_d({\mathbf k}-{\mathbf k}'),\label{disp}
\end{eqnarray}
where $\widetilde V_d({\mathbf k})=c_d\frac{4\pi}{3}\left(\frac{3k_z^2}{k^2}-1\right)$ is the Fourier transform of $V_d({\mathbf r})$ and the last two terms represent the mean-field potentials originating from Hartree direct and Fock exchange interactions, respectively. We remark that the local density approximation has been employed to obtain Eq. (\ref{dist}) for the trapped system.

\begin{figure}
\centering
\includegraphics[width=3.2in]{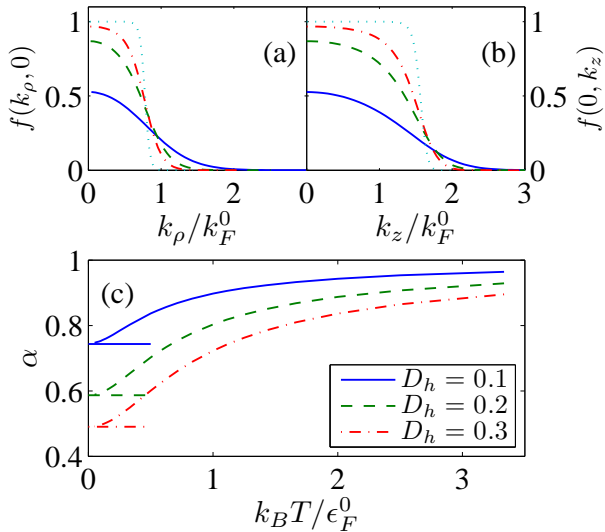}
\caption{(Color online). (a) and (b) represent, respectively, the phase-space distribution functions $f(k_\rho,0)$ and $f(0,k_z)$ for $D_h=0.3$. The corresponding temperatures are $k_BT/\epsilon_F^0=1$ (solid lines), $0.5$ (dashed lines), $1/3$ (dash-dotted lines), and $0.1$ (dotted line). (c) The temperature dependence of $\alpha$ for various dipolar interaction strengths. The horizontal lines denote the values of $\alpha$ at zero temperature.}
\label{hdist}
\end{figure}

Equations (\ref{dist})-(\ref{disp}) form a closed system of equations which can be solved numerically through an iterative procedure to obtain the phase-space distribution function $f({\mathbf r},{\mathbf k})$. Within semi-classical approximation, the total energy at the finite temperature takes the same form as that at zero temperature, i.e., 
\begin{eqnarray}
E\!\!&=&\!\!\frac{1}{(2\pi)^3}\int d{\mathbf r}d{\mathbf k}\left[
\frac{\hbar^2k^2}{2m}+U_{\mathrm{ext}}(\mathbf{r})\right]f(\mathbf{r},\mathbf{k})\nonumber\\
&&\!\!+\frac{1}{2(2\pi)^6}\int d{\mathbf r}d{\mathbf k}
d{\mathbf r}'d{\mathbf k}'f(\mathbf{r},\mathbf{k}) f(\mathbf{r}',\mathbf{k}')V_{d}(\mathbf{r}-\mathbf{r}')\nonumber\\
&&\!\!-\frac{1}{2(2\pi)^6}\int d{\mathbf r}d{\mathbf k}
d{\mathbf k}'
f(\mathbf{r},\mathbf{k})f(\mathbf{r},\mathbf{k}')\widetilde V_{d}({\mathbf k}-{\mathbf k}'),\nonumber
\end{eqnarray}
where the first line represents the kinetic ($E_{\rm kin}$) and potential ($E_{\rm pot}$) energies, and last two lines are, respectively, the direct ($E_{\rm dir}$) and exchange ($E_{\rm exc}$) interaction energies. Finally, we point out that, for axially symmetric system, the phase-space distribution function reduces to $f({\mathbf r},{\mathbf k})=f(\rho,z,k_\rho,k_z)$, where $\rho=\sqrt{x^2+y^2}$ and $k_\rho=\sqrt{k_x^2+k_y^2}$. This fact can be used to simplify the numerical integration~\cite{zhang}.

\section{results}\label{resu}
In this section, we present our results on the thermodynamic properties of a dipolar Fermi gas. For a homogeneous gas, we first obtain the phase-space distribution function numerically, which allows us to calculate other thermodynamic quantities, such as pressure, entropy, and heat capacity. At low temperature limit and in weak interaction regime, we also derive an analytic expression for the entropy, which qualitatively agrees with our numerical result. In the second part of this section, we present the numerical results on the thermodynamics of the trapped systems.

\subsection{Homogeneous case}
\begin{figure}
\centering
\includegraphics[width=2.7in]{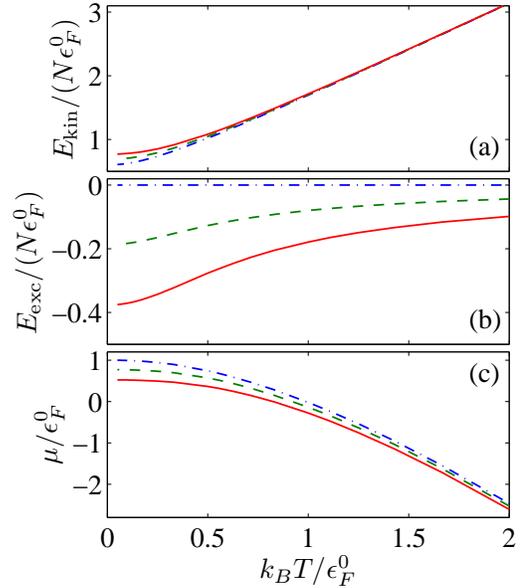}
\caption{(Color online). The temperature dependence of $E_{\rm kin}$ (a), $E_{\rm exc}$ (b), and $\mu$ (c). The dimensionless dipolar interaction strength are $D_h=0$ (dash-dotted lines), $0.2$ (dashed lines), and $0.3$ (solid lines).}
\label{henergy}
\end{figure}

For a homogeneous gas, the phase-space distribution function reduces to a function of momentum, i.e., $f({\mathbf r},{\mathbf k})=f({\mathbf k})$. To present our results, it is convenient to introduce a dimensionless dipolar interaction strength $$D_h=\frac{nc_d}{\epsilon_F^0},$$ where $n=N/{\cal V}$ is the number density of the system and $\epsilon_F^0=\hbar^2(k_F^0)^2/(2m)$ with $k_F^0=(6\pi^2n)^{1/3}$. We note that $k_F^0$ and $\epsilon_F^0$ are, respectively, the Fermi wavevector and Fermi energy of an ideal Fermi gas. 

We plot the phase-space distribution function $f({\mathbf k})$ in Fig. \ref{hdist} (a) and (b) for various temperatures and $D_{h}=0.3$. At low temperature, the momentum distribution is clearly stretched along $z$-axis. However, as one increases the temperature, the momentum distribution becomes less anisotropic. Such behavior can be most easily visualized by calculating the aspect ratio of the cloud in momentum space $\alpha\equiv\sqrt{\langle k_x^2\rangle/\langle k_z^2\rangle}$. Figure \ref{hdist} (c) shows the temperature dependence of $\alpha$ corresponding to various dipolar interaction strengths. For small $T$, $\alpha$ approaches to the value at zero temperature; as one increases the temperature, $\alpha$ goes to unit asymptotically.

\begin{figure}
\centering
\includegraphics[width=2.7in]{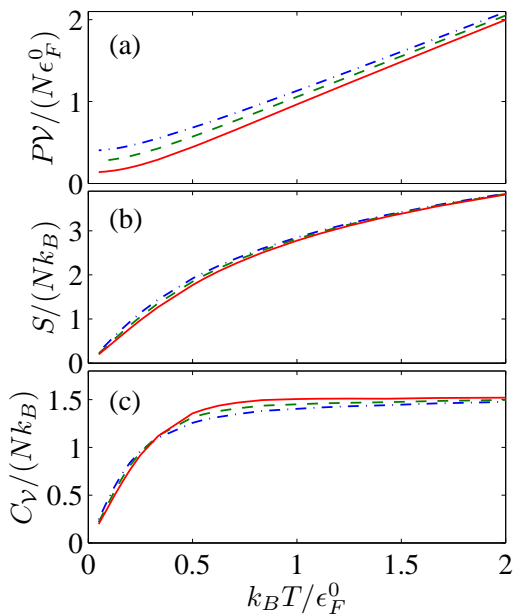}
\caption{(Color online). The temperature dependence of the pressure (a), entropy (b), and heat capacity (c). The dimensionless dipolar interaction strength are $D_h=0$ (dash-dotted lines), $0.2$ (dashed lines), and $0.3$ (solid lines).}
\label{hpcv}
\end{figure}

Due to the spatial homogeneity of the system, the direct dipolar interaction energy vanishes. Therefore, the total energy only contains the contributions from kinetic energy and Fock exchange interaction. In Fig.~\ref{henergy} (a) and (b), we plot, respectively,  $E_{\rm kin}$ and $E_{\rm exc}$ as functions of the temperature. The kinetic energy corresponding to stronger dipolar interaction is larger than that with weaker dipolar interaction, since the stronger dipolar interaction results in larger momentum space deformation. As to the exchange interaction energy, we see that $E_{\rm exc}$ vanishes as $\alpha$ approaches $1$ at high temperature limit. We also plot the temperature dependence of the chemical potential in Fig.~\ref{henergy} (c). Clearly, $\mu$ decreases as one increases $T$, in analogy to the ideal Fermi gas. In addition, the chemical potential is also a decreasing function of $D_h$. This can be most easily understood at zero temperature limit, where the chemical potential becomes~\cite{diff}
\begin{eqnarray}
\frac{\mu_0}{\epsilon_F^0}\simeq \left(1-\frac{8\pi^2}{15}D_h^2\right)^{2/3}\label{chem}
\end{eqnarray}
in weak interaction regime ($D_h\ll1$). Equation (\ref{chem}) indicates that, for given density $n$, the chemical potential decreases as the dipolar interaction strength grows. 

To calculate other thermodynamic quantities, we work with the thermodynamic potential
\begin{eqnarray}
\Omega(T,{\cal V},\mu)=E-TS-\mu N,
\end{eqnarray}
where
\begin{eqnarray}
\frac{S}{{\cal V}k_B}\!=\!-\!\int\!\! \frac{d{\mathbf k}}{(2\pi)^3}\Big\{f({\mathbf k})\ln f({\mathbf k})+[1-f({\mathbf k})]\ln[1-f({\mathbf k})]\Big\}.\nonumber
\end{eqnarray}
is the entropy of the system. The pressure directly relates to the thermodynamics potential as $P{\cal V}=-\Omega(T,{\cal V},\mu)$ and the heat capacity can also be evaluated using the definition $C_{\cal V}=T\left(\frac{\partial S}{\partial T}\right)_{{\cal V}N}$. In Fig.~\ref{hpcv}, we present the temperature dependence of $P$, $S$, and $C_{\cal V}$ corresponding to different dipolar interaction strengths. As a comparison, we also plot the results for an ideal Fermi gas. In general, the results for dipolar Fermi gases approach to those corresponding to the ideal gas at high temperature limit. 

Figure~\ref{hpcv} (a) also indicates that, at a given temperature, $P$ is a decreasing function of $D_h$, in agreement with the zero temperature result~\cite{sogo}. This can be understood since the overall dipolar interaction is attractive for homogeneous gas. In addition, even though $S$ vanishes at zero temperature, the entropy for given $T\neq 0$ also exhibits a similar dependence on $D_h$ as the pressure [Fig.~\ref{hpcv} (b)]. Furthermore, The temperature and interaction dependence of the heat capacity, shown in Fig.~\ref{hpcv} (c), can be naturally deduced from the behavior of th entropy.

To gain more insight into the entropy, we shall derive an analytical expression for $S$ at low temperature limit. To this end, we note that when $T\rightarrow 0$, the derivative of $S$ with respect to $T$ at constant $\cal V$ and $\mu$ can be approximately expressed as~\cite{fetter}
\begin{eqnarray}
T\left(\frac{\partial S}{\partial T}\right)_{{\mathcal V}
\mu}={\cal V}\int\frac{d{\mathbf
k}}{(2\pi)^3}\left(\varepsilon_0({\mathbf k})-\mu\right)\frac{\partial f({\mathbf k},T)}{\partial T},\nonumber
\end{eqnarray}
where we have explicitly expressed $f$ as a function of the temperature $T$ and the zero temperature dispersion relation for the quasi-particle is~\cite{chan}
\begin{eqnarray*}
\varepsilon_0({\mathbf k})\simeq \frac{\hbar^2k^2}{2m}-2\epsilon_F^0D_hP_2(\cos\theta)I\left(\frac{k}{k_F^0}\right),
\end{eqnarray*}
with $P_2(\cdot)$ is the second order Legendre function, $\theta$ is the polar angle of $\mathbf k$, and
\begin{eqnarray*}
I(x)=\frac{\pi}{12}\left\{3x^2+8-\frac{3}{x^2} +\frac{3\left(1-x^2\right)^3}{2x^3}{\rm
ln}\left\vert\frac{1+x}{1-x}\right\vert\right\}.
\end{eqnarray*}
Following the same procedure as that in Ref.~\cite{fetter}, we find
\begin{eqnarray}
S(T, {\mathcal V}, \mu)\simeq T\left(\frac{\partial S}{\partial T}\right)_{{\mathcal V}
\mu}\simeq \frac{\cal V}{12}k_B^2TA(\mu,c_d),\label{entropy}
\end{eqnarray}
where
\begin{eqnarray}
A(\mu,c_d)=\int\left[k^2\frac{dk}{d\varepsilon_0(k,\theta)} \right]_{\varepsilon_0(k,\theta) =\mu}\sin\theta d\theta.\nonumber
\end{eqnarray}
We note that Eq. (\ref{entropy}) is not the entropy in usual sense as it is a function of $\mu$. To proceed further, we change variable from $\mu$ to $N$ by using Eq. (\ref{chem}), which yields, at low temperature limit and in weak interaction regime,
\begin{eqnarray}
\frac{S(T,{\cal V},N)}{Nk_B}\simeq\frac{\pi^2}{2}\frac{k_BT}{\epsilon_F^0} \left(1-\frac{8\pi^2}{15}D_h^2\right)^{1/3}\left(1+\frac{7\pi^2}{45}D_h^2\right). \nonumber
\end{eqnarray}
This expression clearly indicates that the entropy is decreasing function of $D_h$, in qualitative agreement with our numerical results.

\subsection{Trapped case}

Now we turn to study the trapped case, for which we shall adopt a set of dimensionless units based on the harmonic oscillator length $\bar a=\sqrt{\hbar/(m\overline\omega)}$: $N^{1/6}\bar a$ for length, $N^{1/6}\bar a^{-1}$ for wave vector, and $N^{1/3}\hbar\overline\omega$ for energy. Under these choices, we may define a dimensionless dipolar interaction strength $$D_t=\frac{N^{1/6}c_d}{\hbar\overline\omega \bar a^3}.$$ In addition, the normalization condition for phase-space distribution function becomes $(2\pi)^{-3}\int d{\mathbf r}d{\mathbf k}f({\mathbf r},{\mathbf k})=1$.

\begin{figure}
\centering
\includegraphics[width=3.4in]{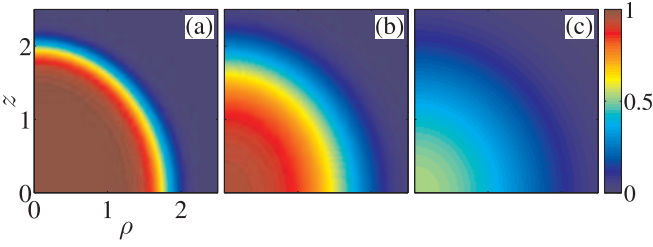}
\caption{(Color online). The phase-space distribution function $f(\rho,z,0,0)$ of a trapped gas in spherical potential for $k_BT/(N^{1/3}\hbar\bar\omega)=0.2$ (a), $0.5$ (b), and $1$ (c). The dimensionless dipolar interaction strength is $D_t=1$.}
\label{tdist}
\end{figure}
\begin{figure}
\centering
\includegraphics[width=2.7in]{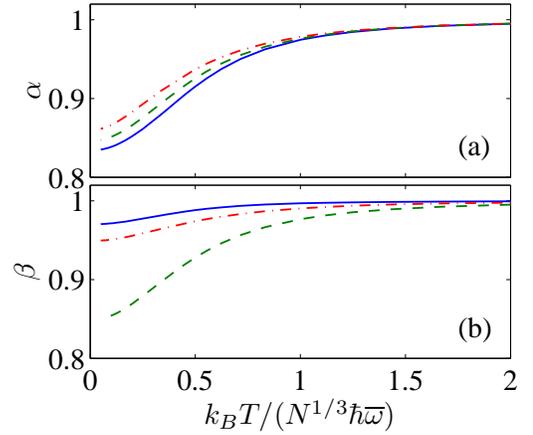}
\caption{(Color online). The temperature dependence of the deformation parameters $\alpha$ (a) and $\beta$ (b) for $\lambda=0.1$ (solid lines), $1$ (dashed lines), and $10$ (dash-dotted lines). The dimensionless dipolar interaction strength are $D_t=1$.}
\label{tdefor}
\end{figure}

\begin{figure}
\centering
\includegraphics[width=2.7in]{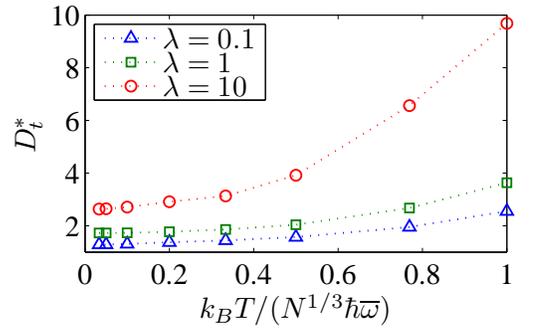}
\caption{(Color online). The temperature dependence of the critical dipolar interaction strengths for various trap geometries.}
\label{tcri}
\end{figure}

In Fig.~\ref{tdist}, we present the typical results for phase-space distribution function $f(\rho,z,0,0)$ for a gas trapped in a spherical potential. Similar to the momentum space distribution for a homogeneous gas (Fig.~\ref{hdist}), the distribution function in real space becomes elongated along $z$-axis at low temperature. However, the deformation becomes smaller and smaller as the temperature is increased. To characterize the real space deformation, we define the deformation parameter in real space as $\beta\equiv\lambda^{-1}\sqrt{\langle x^2\rangle/\langle z^2\rangle}$, where the inclusion of the factor $\lambda^{-1}$ is to eliminate the deformation caused by trapping potential such that $\beta=1$ for a noninteracting gas. Figure~\ref{tdefor} (a) and (b) plot, respectively, the temperature dependence of the deformation parameters in momentum and real spaces. As expected, stronger dipolar interaction induces larger deformations in momentum and real spaces. However, both $\alpha$ and $\beta$ approach unit at high temperature limit.

Finally, we explore the stability of the trapped gases at finite temperature. Similar to the zero temperature case~\cite{miya,zhang}, due to the partially attractive feature of the dipolar interaction, there also exists a critical dipolar interaction strength $D_t^*$ for a given temperature $T$, beyond which the system becomes unstable. In Fig.~\ref{tcri}, we plot the temperature dependence of the critical dipolar interaction strength for various trap geometries. It can be seen that, independent of the trap aspect ratio $\lambda$, $D_t^*$ is an increasing function of $T$. This can be understood as higher temperature corresponds to smaller momentum and real space deformations, resulting in smaller dipolar interaction. In addition, higher temperature also yields higher kinetic energy which stabilizes the system.

\section{Conclusions}\label{concl}
To conclude, based on the semi-classical theory, we have studied the thermodynamics of a dipolar Fermi gas by numerically finding the phase-space distribution function. We show that the deformations in both momentum and real space becomes smaller and smaller as one increases the temperature. For homogeneous gases, we also calculate pressure, entropy, and heat capacity. In particular, at low temperature limit and in weak interaction regime, we obtain an analytic expression for the entropy, which agrees qualitatively with our numerical result. The stability diagram for a trapped gas at finite temperature is also obtained.

This work was supported by NSFC through grants 10974209 and 10935010, by the National 973 program (Grant No. 2006CB921205), and by the ``Bairen" program of Chinese Academy of Sciences.


\begin{thebibliography}{}
\bibitem{diff} Equation (\ref{chem}) is different from that obtained in Ref.~\cite{ronen}.
\end{thebibliography}
\end{document}